\documentclass[11pt]{article}
\usepackage[utf8]{inputenc}
\usepackage[T1]{fontenc}
\usepackage{lmodern}
\usepackage{amsmath}
\usepackage{amsfonts}
\usepackage{mathtools}
\usepackage{amssymb}
\usepackage{amsthm}
\usepackage{color}
\usepackage{xcolor}
\usepackage[pdftex]{graphicx}
\usepackage{hyperref}
\usepackage{cleveref}
\usepackage{csquotes}
\usepackage{mathrsfs}
\usepackage{stmaryrd}
\usepackage[english]{babel}
\usepackage{xspace}
\usepackage{etoolbox}
\usepackage{tikz-cd}
\theoremstyle{plain}
\newtheorem{definition}{Definition}[section]

\newtheorem{theorem}[definition]{Theorem}

\newtheorem{lemma}[definition]{Lemma}
\newtheorem{proposition}[definition]{Proposition}
\newtheorem{remark}[definition]{Remark}
\def\rmEl{{\rm {El}}}

\def\QQ{\mathbb{Q}}
\def\RR{\mathbb{R}}
\begin{document}

\title{A computation of maximum likelihood for 4-states-triplets under Jukes-Cantor and MC }
\author{
  author1\\
 \texttt{mariemi1@ucm.es}
  \and
 author2\\
  \texttt{eralva01@ucm.es}}

\author{Mar\'{\i}a Emilia Alonso $^1$\thanks{Partially supported by,
  PID2020-114750GB-C32, PID2024-156181NB-C31, by GRUPO SINGULAR UCM and IMI(UCM).} 
\and Ernesto \'Alvarez $^2$\thanks{Partially supported by,
  PID2020-114750GB-C32.} }
  
 \date{
$^1${ Depto. de \'Algebra, Geometr\'{\i}a y Topolog\'{\i}a, UCM, 28040 Madrid, Spain, 
{\tt mariemi@mat.ucm.es}.}\\
$^2${Univ. UABJO, Oaxaca, M\'exico, \\ {\tt eralav01@ucm.es}.}
}

\maketitle


\begin{abstract}
We study the \cite{ChorHendySnir2006} evolutionary model, which consists of a rooted phylogenetic tree  with three leaves, subject to the Jukes--Cantor (JC69) molecular evolutionary model and molecular clock. We show that the  likelihood function associated with this model has a unique maximum which depends analytically of the parameters (as it was conjectured in \cite{ChorHendySnir2006}), assuming that these parameters verify some very precise inequalities; some of which arise naturally from the model.
With a typical argument of differential topology we reduce the proof to answer 
 a question of algebra, very simple, although computationally involved, that we solve using  some Maple libraries.
 We are very indebted to Marta Casanellas, who presented the problem to us and gave us the first insights on it.
\end{abstract}

 
 
 \section{The 4-states Triplets under Jukes-Cantor model of evolution, following the approach of Chor-Hendy-Snir (2006)}
  
Phylogenetic reconstruction consists of determining the most probable ancestral relationships that gave rise to current species (\cite{Felsenstein2004, Casanellas2018}). These species can be identified by their nucleotide sequences (adenine, guanine, cytosine, and thymine), which we refer to in this paper as “characters” (or states). 

The starting point (\cite{Felsenstein2004}), that  provides the only known information, is the so called alignment. That consists in identifying homologous character sequences of extant species that are related (ie. that share a common ancestor). This information is organized  in such a way that each column of the alignment correspond to a character pattern (within this context, we refer to the columns as sites). The normalized frequencies of the appearances of the different characters will be called observed data (or the observed probabilities). We also assume that the different sites of the alignement evolve independently and are identically distributed.

The second ingredient is the consideration of a concrete ``evolution model'', that consists of a phylogenetic tree (connected, acyclic, binary graph eventually with a node playing the role of a root), in which each vertex corresponds to a discrete random variable, whose states are nucleotides. The random variables of the vertices of degree 1 (known as leaves) are observed (since they correspond to current species, whose genetic code is known), while the random variables of the interior vertices are hidden (since they correspond to ancestral species). 
 We associate these evolutionary models with a continuous-time Markov chain structure, defining transition matrices on each of their edges, which estimate the probabilities of state change between contiguous vertices. As an example in the Figure \ref{tripode1} observe a rooted phylogenetic tree with three leaves whose edges are labelled by $e_i$'s.

An excellent book about the algebraic techniques used in phylogenetics (as the so called  evolutionary model for 4 states Jukes--Cantor considered here and other graphical models) is the book of Sullivan\cite{Sullivant2010}. However we will  only use here our main reference \cite{ChorHendySnir2006}. 
Indeed, our study begins with the evolutionary model of \cite{ChorHendySnir2006}, which is a rooted phylogenetic tree with three leaves (see Figure \ref{tripode1}), subject to the Jukes-Cantor (JC69) molecular evolutionary model; which involves for every edge $e$ the presence of symmetric $4\times 4$ transition matrices with the same structure, a single free parameter $q_e$ and identical terms on the diagonal, such as the matrix $P_{q_e}$ displayed in  (\ref{matrizjc69}).

{\small

\begin{equation}\label{matrizjc69}
P_{q} =\frac{1}{4}
\bordermatrix 
{
	~ & A & C & G & T \cr
	A & 1+3\exp(-4q) & 1-\exp(-4q) & 1-\exp(-4q) & 1-\exp(-4q) \cr
	C & 1-\exp(-4q) & 1+3\exp(-4q) & 1-\exp(-4q) & 1-\exp(-4q) \cr
	G & 1-\exp(-4q) & 1-\exp(-4q) & 1+3\exp(-4q) & 1-\exp(-4q) \cr
	T & 1-\exp(-4q) & 1-\exp(-4q) & 1-\exp(-4q) & 1+3\exp(-4q)
},    
\end{equation}
}

\smallskip

\begin{figure}[htb]
	\centering
	\setlength{\unitlength}{0.15mm}
	\begin{picture}(350,100)
	{
		\tiny 
		\put(150,100){\vector(-2,-3){50}}
		\put(100,25){\vector(-2,-3){50}}
		\put(150,100){\vector(2,-3){100}}
		\put(100,25){\vector(2,-3){50}}
		\put(45,-59){$1$}
		\put(150,-59){$2$}
		\put(250,-59){$3$}
				\put(10,-27){{\tiny $P_{q_{e_1}}$}}
		 		\put(140,-27){{\tiny $P_{q_{e_2}}$}}
				\put(72,65){{\tiny $P_{q_{e_3}}$}}
		\put(92,28){$I$}
		\put(144,103){$R$}
	}
	\end{picture}
		\bigskip
	\caption{Phylogenetic tree model by  \cite{ChorHendySnir2006}.}
\label{tripode1}
\end{figure}
\smallskip

\noindent According to this model, the maximum number of character patterns observed within an alignment in the tripod\ref{tripode1} is $4^3$. However the authors in \cite{ChorHendySnir2006} manage to represent these character patterns in $16$ \textit{substitution patterns}. A substitution pattern for an alignment of three species is an ordered pair, $({\sf X},{\sf Y})$, with ${\sf X}, {\sf Y}\subset\{1,2\}$ such that
$i\in {\sf X}$ if and only if  the transformation required to pass from the nucleotide observed in the third leaf to the one in the  $i$-leaf is a $\alpha$ or $\gamma$ transformation, $i\in {\sf Y}$ if and only if the transformation required to pass from the nucleotide observed in the third leaf to the leaf $i$ is a $\gamma$ or $\beta$ transformation (see the diagram below).

\[
\begin{tikzcd}[row sep=6em,column sep=6em]
A \arrow{r}{\alpha} \arrow{d}{\beta} \arrow{dr}{\gamma}& G\arrow{l}[swap]{\alpha}\arrow{d}{\beta}\arrow{dl}[swap]{\gamma}  \\
 T \arrow{r}{\alpha} \arrow{u}[swap]{\beta}\arrow{ur}{\gamma} & C\arrow{l}[swap]{\alpha}\arrow{u}[swap]{\beta}\arrow{ul}[swap]{}
\end{tikzcd}
\]

\smallskip

\noindent So that  $(\{ 1 \},\{ 1,2 \})$ means that, to transform the observed nucleotide in the first leaf to the one observed in the third leaf we need  a $\gamma$ transformation (see below)  and from the second leaf to the third, we need a $beta$ transformation. So that, if in the third leaf there is  $A$, there is  $C$ in the first leaf and $T$ in the second. 
In  the same way $(\{ \emptyset \},\{ 1\})$ means that to transform the observed nucleotide in the first leaf to the one observed in the third leaf we need  a $\beta$ transformation and in the second leaf there is the same character that in the third. So that, if in the third leaf there is  $A$, there is a $T$ in the first and $A$ in the second.
 The rest of substitution patterns that tell us the nucleotide in second leaf is the same as the one in the third, but the one in the first and in the third are different are  $(\{ 1 \},\{ \emptyset\}), (\{ 1\}, \{ 1\})$. So we collect the frequencies of the substitution patterns as follows:
\begin{align*}
f_0 = f_{(\emptyset,\emptyset)},\\
f_1 = f_{(\emptyset,\{1\})}+f_{(\{1\},\emptyset)}+f_{(\{1\},\{1\})},\\
f_2 = f_{(\emptyset,\{2\})}+f_{(\{2\},\emptyset)}+f_{(\{2\},\{2\})},\\
f_3 = f_{(\emptyset,\{1,2\})}+f_{(\{1,2\},\emptyset)}+f_{(\{1,2\},\{1,2\})},\\
f_4 = f_{(\{1\},\{2\})} + f_{(\{1\},\{1,2\})} + f_{(\{2\},\{1\})} + f_{(\{2\},\{1,2\})} + f_{(\{1,2\},\{1\})} + f_{(\{1,2\},\{2\})}.
\end{align*}

\noindent One can see that $f_1$ (resp $f_2$) represents the frequency of the substitution patterns that have in the second leaf (resp. in the first leaf ) the same nucleotide as in the third, and different from the one observed in the first (resp. in  the second  leaf), and $f_3$ is the frequency of the substitution patterns having the same nucleotide in the first and second leaf and different from the one in the third leaf.

The usual prune algorithm to compute the probabilities of  the occurrence of three given characters $(i,j,k)$ at the leaves, in terms of
 the $q_i$'s
 , can be adapted to estimate the probabilities of the substitution patterns in this model . In \cite{ChorHendySnir2006} the authors apply \textit{ Hadamard conjugation} to this tree and the evolution to calculate the probability distribution of the 16 associated substitution patterns in terms of $q_i$ (see \autoref{tripode1}). After transforming the variables $x_i := e^{-4q_i}$ in terms of $x_i$'s they get,

\[ P =
\bordermatrix
{
	~ & \emptyset & \{1\} & \{2\} & \{1,2\} \cr
	\emptyset &  {\sf a_0} &  {\sf a_1} &  {\sf a_2} &  {\sf a_3} \cr
	\{1\} &  {\sf a_1} &  {\sf a_1} &  {\sf a_4} &  {\sf a_4} \cr
	\{2\} &  {\sf a_2 }&  {\sf a_4} &  {\sf a_2} &  {\sf a_4 }\cr
	\{1,2\} &  {\sf a_3} &  {\sf a_4} &  {\sf a_4} &  {\sf a_3}
},
\]

where

\begin{equation}\label{modelo estadistico de hendy}
\begin{pmatrix}
 {\sf a_0} \cr
 {\sf a_1} \cr
 {\sf a_2} \cr
 {\sf a_3} \cr
 {\sf a_4}  
\end{pmatrix}
=
\frac{1}{16} 
\begin{pmatrix}
1 & 3 & 3 & 3& 6 \cr
1 & 3 & -1 & -1& -2 \cr
1 & -1& 3  &  -1   &   -2 \cr
1 &-1 & -1 & -1 &  2    
\end{pmatrix}
\begin{pmatrix}
1 \cr
x_2x_3 \cr
x_1x_3 \cr
x_1x_2  \cr
x_1 x_2 x_3
\end{pmatrix}.   
\end{equation}

\noindent 
Finally in \cite{ChorHendySnir2006}, the authors assume a big simplification adding the hypothesis of 
{\it molecular clock } to the evolution model;  the existence of a single mutation rate $\lambda$ that produces every transition matrix $P_{q_e}$ along 
the phylogenetic tree. That means that $q_e = \lambda t_e$, and thus $q_e$  stands for the expected value of mutations after time $t_e$ has elapsed. Hence in our case of the rooted tripod \autoref{tripode1} we have that $q_1=q_2 \leq q_3$, and therefore $ x_1=x_2\geq x_3$ and $ {\sf a_1}= {\sf a_2}$. 

 \noindent B. Chor, et alt. in \cite{ChorHendySnir2006} raised the problem of the unicity of the maximum of the likelihood function associated to this model; giving some examples. 
 
 As it is well known, given the observed probabilities of the substitution patterns, the maximum(s) of the likelihood function provide us with optimal values of the parameters of the phylogenetic tree model already described. 
 The likelihood function obtained of this model is:
 
\begin{equation}\label{}
 L( {\sf a},f) =  {\sf a_0}^{f_0} {\sf a_1}^{f_1} {\sf a_2}^{f_2} {\sf a_3}^{f_3} {\sf a_4}^{f_4}    
\end{equation}

\noindent where the $f_i$'s are the observed frequencies of the substitution patterns defined as before.
Under the hypothesis of molecular clock, setting $f_{12}= f_1+f_2$ the function to maximize becomes: 

$$ L( {\sf a},f) =  {\sf a_0}^{f_0} {\sf a_2}^{f_{12}} {\sf a_3}^{f_3} {\sf a_4}^{f_4} $$  

We show in this paper that the likelihood function of this model, under the hypothesis of molecular clock has a unique maximum which depends analytically of the parameters, assuming that these parameters verify very precise inequalities, some of which arise naturally from the model.

\bigskip
 
\noindent {\bf Remark.} We remark that the variables are $x_2$ and $x_3$, and since we have fixed the phylogenetic three as in \autoref{tripode1},
 the molecular clock give $t_3\geq t_1=t_2$ or, equivalently, $0\leq x_3 \leq x_2=x_1$. 
The  frequencies of the observed data are $f_0, f_{12}, f_3, f_4$ and they will be called {\sl parameters}.
 Notice also that for  $x_2,x_3$ in the region ${\mathcal A}:=\{ (x_2,x_3): 0\leq x_3 \leq x_2\leq 1 \}$, the probabilities of the model, that is the  $a_i's \in [0,1]$ verify some inequalities. Indeed we have:
 $ {\sf a_0}- {\sf a_3} = x_2x_3(1+x_2)/2\geq 0$, $ {\sf a_3}-  {\sf a_{12}}= x_2(x_2-x_3)/4\geq 0$, $ {\sf a_{12}}- {\sf a_4}= x_2x_3(1-x_2)/4 \geq 0$, and ${\sf a_4}=(1-x_2)(1-x_2x_3 + x_2(1-x_3))/16 \geq 0$ for  $0\leq x_3 \leq x_2\leq 1$.  
 Moreover $ {\sf a_0} + 3 {\sf a_3} + 6 {\sf a_{12}} + 6 {\sf a_4}=1$ because it is the sum of all probabiities of the substitution patterns.
 
 \noindent Therefore in the following we shall assume that these inequalities are also fulfilled by the observed data or, in other words
 $f_0\geq f_3 \geq f_{12} \geq f_4 \geq 0$, and that they are normalized so that $1=f_0 + 3f_3 + 6f_{12} + 6f_4$. The semi-algebraic set determined by all these inequalities of the parameters will be called {\sl admissible region for the parameters}.

 \bigskip
 \section{Uniqueness of the ML and analytical dependency on the parameters}
 
\noindent The objective of our study is to answer the question in \cite{ChorHendySnir2006}: {\it is there a unique maximum for $L({\sf a},f)$ (or equivalently for $h_f:= \ln L({\sf a},f)$), for a given value of the parameters in the admissible region, with analytical dependence on the parameters?}

The first naive attempt to 
try to solve the equations defining the critical points of $h_f$, (namely, find solutions in the interior of ${\mathcal A}$ where the two partial derivatives of $h_f$ vanish),  leads to an intractable system of equations,  
that caused to collapse all the software of Groebner basis we know \cite{Alvarez2025}. Moreover our problem requires elimination over the real numbers, so we will use a bit of differential topology to turn this into a first-year algebra exercise, but so computationally involved that it requires the use of Maple libraries.

We will follow  \cite{Christensen2017}, who provides sufficient conditions on a real-valued function to ensure the existence of a unique global maximum: 

\begin{theorem}\label{christensen} (\cite{Christensen2017}, here for $n=2$)
 \noindent Let $h:M\subset \mathbb{R}^2\rightarrow \mathbb{R}$ be a Morse function defined on a smooth, compact, contractible manifold. If $M$ has a boundary, suppose that the gradient of $h$, $\nabla h$ is well defined and points inward from its boundary. Then $h$ has a unique critical point in $M$, which is a global maximum, if and only if $\nabla h(x^*) = 0 \implies det(D^2h(x)) > 0$. 
\end{theorem}

This theorem is a direct consequence of Poincar\'e-Hopf theorem and Morse lemma. We are going to apply it to $h_f$.

\begin{multline}\label{hf}
h_f := \ln (L( {\sf a},f)) = (1-3f_3-6f_{12}-6f_4) \ \ln \ (6x_2^2x_3+3x_2^2+6x_2x_3+1)+ \\
f_{12} \ \ln \ (-2x_2^2x_3-x_2^2+2x_2x_3+1)+f_3 \ \ln \ (-2x_2^2x_3+3x_2^2-2x_2x_3+1)+  \\
f_4\ \ln \ (2x_2^2x_3-x_2^2-2x_2x_3+1)
\end{multline}

\medskip

\noindent \mbox {where we have reduced the number of parameters, using  $f_0+6f_{12}+3f_3+6f_4=1$}.

\noindent Simple tests show that the arguments of the logarithmic functions do not vanish in the interior of the triangle with vertices 
$(0,0), (1,0), (1,1)$ of the plane $x_2, x_3$. Indeed, $2x_2^2x_3-x_2^2-2x_2x_3+1=x_3(x_2-1)^2+(1-x_3)(1-x_2^2)$,  $-2x_2^2x_3-x_2^2+2x_2x_3+1=2x_2^2(1-x_3)+2x_2(x_2-x_3)-x_2^2+1$, and $-2x_2^2x_3-x_2^2+2x_2x_3+1=2x_2x_3(1-x_2)+1-x_2^2$ .

\noindent We define a {\sl restricted  domain} of $h_f$, w.r.t. the variables $x_2,x_3$ , as ${\mathcal A}'= \{(x_2,x_3) | 0\leq x_3\leq x_2 \leq 1-\epsilon\}$ with $\epsilon > 0$  very small number. Clearly ${\mathcal A}'\subset {\mathcal A}$ is contractible, and, in order to apply \cite{Christensen2017}, avoiding the singularities of the corners, you can ``smooth the corners'' by getting as close as you want to the vertices of the triangle. 

\subsection{Description of a suitable region of the parameters}

In order to apply theorem \ref{christensen}, we need to restrict the admissible region of parameters. At the end of the paper we have some comments \ref{notabiologica} about the meaning of this restriction in relation to the localization of the maximum. 

\begin{lemma}
For every value of the parameters $f$'s in the admissible region, verifying in addition $f_3 - f_{12} - \frac{2}{3}f_4 \geq 0$,
the gradient of $h_f$ (see \ref{hf}), $\nabla h_f$, points inward along the boundary of ${\mathcal A}'$.
\end{lemma}

\noindent {\bf Proof}.
 We must verify that on each side of the right triangle ${\mathcal A}'$  the gradient points toward the interior of ${\mathcal A}'$.
 
The partial derivatives $\partial h_f/\partial x_2$ and $\partial h_f/\partial x_3$ share the same denominator 

{\footnotesize{ 
$$D(x_2,x_3)=(6x_2^2x_3+3x_2^2+6x_2x_3+1)(x_2-1)^2(2x_2x_3+x_2+1)(2x_2^2x_3-3x_2^2+2x_2x_3-1)(2x_2x_3-x_2-1)$$
}}

\noindent which is strictly positive on ${\mathcal A}'$. Indeed, the three first factors are positive in ${\mathcal A}'$ while the two last are negative in ${\mathcal A}'$. 
 Let $N_1 = D(x_2,x_3)\partial h_f/\partial x_2$ and $N_2 = D(x_2,x_3)\partial h_f/\partial x_3$ (See the Appendix).
We will call $(N_1, N_2)$ the {\sl reduced gradient}.
We consider the sign of the reduced gradient along the hypotenuse of the triangle ${\mathcal A}'$, $x_2=x_3$. The vector perpendicular to it pointing inward, is $(1,-1)$. The scalar product of the vector with the gradient is $\frac{N_1-N_2}{D(x_2,x_3)}$. Therefore, the sign of the scalar product coincides with the sign of $N_1-N_2$. 

\begin{multline*}
N_1-N_2 = -2x_3(2x_3^2+x_3+1)(x_3-1)^3\\
(60f_{12}x_3^5+24f_3x_3^5+60f_4x_3^5+36f_{12}x_3^4+48f_3x_3^4+48f_4x_3^4-12x_3^5-51f_{12}x_3^3\\
+54f_3x_3^3-33f_4x_3^3-12x_3^4-65f_{12}x_3^2-2f_3x_3^2-65f_4x_3^2+3x_3^3-57f_{12}x_3\\
-22f_3x_3-55f_4x_3+9x_3^2-19f_{12}-6f_3-19f_4+9x_3+3)
\end{multline*}

\noindent Since $ -2x_3(2x_3^2+x_3+1)(x_3-1)^3$ this factor is positive in the region ${\mathcal A}'$, the  sign of $N_1-N_2$ on  the hypotenuse is the same as that of the factor dependent on the parameters $f_{12},f_3$ and $f_4$. Let us call this factor $P(f_{12},f_3,f_4;x_3) = P$. We have the following calculations related to $P$:
$P(x_3=0) = -19f_{12}-6f_3-19f_4+3 = -19f_{12}-6f_3-19f_4+3(f_0+3f_3+6f_{21}+6f_4) = 3f_0+3f_3-f_{12} -f_4$. This evaluation is positive since the parameters are in the admissible region. To see that $P$ is positive in $[0,1]$ for parameters in the admissible region, it is necessary that  $P(x_3=1)>0$ and that $P$ has no zeros in $[0,1]$. Since $P(1)=f_3 - f_{12} - \frac{2}{3}f_4$, we need to add this inequality $P(1)\geq 0$ to our admissible region of parameters, as a necessary condition to be $P$ positive in $[0,1]$.

 Now, we compute the Budan--Fourier sequence of $P$ in the interval $[0,1]$, and  verify that, under the assumption on the parameters, the Budan-Fourier sequence in $x_3=0$ and in $x_3=1$ both present the same number of sign changes.

{\footnotesize {
\[
 \begin{array}{lll}
  P(0)=3-19f_{12}-6f_3-19f_4=3f_0+3f_3-f_{12}-f_4 \geq 0&&  P(1)=96(f_3-f_{12}-(2/3)f_4 \geq 0 \\
  
  P^{'}(0)=9f_0+5f_3-f_4-3f_{12}\geq 0  && P^{'}(1)=8(-9f_0+29f_3-41f_{12}-28f_4) ?0 \\
  
  P^{''}(0)=2(9f_0+25f_3-11f_{12}-11f_4) \geq 0 && P^{''}(1)=-348f_0+332f_3-892f_{12}-640f_4\leq 0  \\ 
  
   P^{'''}(0)=2(9f_0+189f_3-99f_{12}-45f_4)\geq 0 &&  P^{'''}(1)=-990f_0-54f_3-1782f_{12}-1386f_4 \leq 0\\
   
  P^{iv}(0)= 288(-f_0+f_3-3f_{12}-2f_4)\leq 0 && P^{iv}(1)=-2304f_{12} - 1152f_3 - 2016 f_4 - 1728f_0 \leq 0 \\
  
  P^{v}(0)= -1440(f_{12}+f_3+f_4+f_0) \leq0  &&  P^{v}(1)=-1440(f_{12}+f_3+f_4+f_0) \leq 0
 \end{array}
\]}}

\noindent The signs of the given derivatives above are consequence of the inequalities among the parameters: $ 1=f_0+3f_3+6f_{12}+6f_4$, and $f_0\geq f_3 \geq f_{12} \geq f_4 \geq 0$.
Notice that even if the sign in $'?'$ is  positive or negative, being in between of different signs, it does not produce a new change of sign.\\

\noindent Now we deal with the sign of the reduced gradient along the vertical side of triangle ${\mathcal A}'$. The vector perpendicular to it, pointing inward, is $(-1,0)$. The scalar product of the vector with the reduced gradient is $-N_1$. It turns out that $N_1$ vanishes at $x_2=1$ and  $-N_1=-2(1-x_2) H_1(x_2,x_3)$ for some polynomial $H_1$ such that $H_1(x_2=1,x_3)=-32(f_{12}+f_4)(x_3+1)(3x_3+1)(x_3-1)^2$. So that 
$-N_1$ is positive on the vertical side $x_2= 1- \epsilon $.

\noindent Finally we consider the sign of the gradient along the base of the triangle ${\mathcal A}'$ for $x_3 = 0$. The vector perpendicular to it, pointing inward, is $(0,1)$. The scalar product of the vector with the reduced gradient is $N_2$. Therefore, the sign of the scalar product coincides with the sign of $N_2$. We have
$N_2(x_2,x_3=0)=-2x_2(x_2-1)^2(x_2+1)(3x_2^2+1) M$, where  $M$ is a polynomial of degree $2$  in $x_2$,   $M=x_2^2A+x_2B+C$ where $A$, $B$, $C$ are linear in the parameters $f_i$'s. We have $M(x_2=0)=C= -3+10f_3+17f_{12}+19f_4=-3(f_0+6f_{12}+3f_3+6f_4)+10f_3+17f_{12}+19f_4=-3f_0+f_3-f_{12}+f_4\leq0$. 
Moreover, computing the discriminant of $M$ we get $ 4(f_{12}-f_4)(-12f_0+4f_3-3f_{12}+3f_4)\leq 0$, 
so that $M$ is negative for $x_2 \in [0,1]$ (and the parameters in the admissible region), therefore
$N_2(x_2,x_3=0)$ is positive in ${\mathcal A}'$.

As a consequence of the lemma we introduce the following 

\begin{definition}\label{region}
We call  {\it restricted admissible region of the parameters} to the region: 
\begin{multline}
 1=f_0+3f_3+6f_{12}+6f_4, \ f0\geq f_3\geq f_{12} \geq f_4\geq 0, \ f_3\geq f_{12} +2/3 f_4
\end{multline}

\end{definition}

\subsection{Study of the sign of the determinant of the Hessian of $h_f$ in the region of interest and admissible parameters}

\noindent We know already that $h_f$ increases going through the boundary to the interior of ${\mathcal A}'$, so that the maximum (s) in the compact region ${\mathcal A}'$ is (are) local. We prove now that the second condition of \cite{Christensen2017} is fulfilled. As a consequence, we conclude that there is a unique critical point in the interior of ${\mathcal A}'$, that must be the global maximum. Precisely:

\begin{theorem}
 
The Hessian of $h_f$ (see \ref{hf}) does not vanish at any critical point in (the interior of) ${\mathcal A}'$. Moreover its sign at every critical point of $h_f$ in  ${\mathcal A}'$ is positive for the parameters taking values in the restricted admissible region (see \ref{region}). In particular the ML problem for a tripod under JC69
 evolution model and MC, as stated in \cite{ChorHendySnir2006}, has a unique solution and it depends analytically on the parameters that take values in the polyhedral region \ref{region}. 
\end{theorem}

\bigskip
\noindent {\bf Proof.} 

\noindent The gradient of $h_f$ is $\frac{1}{D(x_2,x_3)}(N_1,N_2)$, with $D$ positive in ${\mathcal A}'$.
Hence, the sign of the determinant of the Hessian of $h_f$ on its critical points  in ${\mathcal A}'$ is the same as the sign  of 
 $N_{11}N_{22}-N_{12}N_{21}$, where $N_ {ij} = \partial N_i/\partial x_{j+1}$, $i,j\in\{1,2\}$ (see the Appendix). Let us call \textit{Hess} to the polynomial $N_{11}N_{22}-N_{12}N_{21}$.
 
 \noindent {\sc The first step of the proof} consists in reducing our sign determination problem of the determinant of the Hessian,  \textit{Hess} on the critical points of $h_f$ under the linear constraints of the parameters \ref{region}, to a simpler problem of real elimination with only one parameter under a quadratic restriction.
 
  It turns out that $N_1$ has a factor equal to $-2(x_2-1)$ 
 and $N_2$ has a factor equal to $-2x_2(x_2-1)^2$. Therefore, $-4(x_2-1)^2$ divides $N_{11}N_{22}-N_{12}N_{21}$, so that the sign of $hessred := -\frac{Hess}{4(x_2-1)^2}$ is the opposite of the sign of \textit{Hess} and hence the opposite of the sign of the determinant  of the Hessian of $h_f$. These given factors  of $N_1$ and $N_2$  vanish only in $x_2=1$ 
 which does not belong to our reduced region ${\mathcal A}'$, and in $(0,0)$ that we have avoided ``smoothing the corners''. So that setting $Ec_1:= \frac{N_1}{-2(x_2-1)}$ and $Ec_2:=\frac{N_2}{-2x_2(x_2 - 1)^2}$,
 we have to prove that  $N_{11}N_{22}-N_{12}N_{21}$ is positive on the set $\{ Ec_1=0, Ec_2=0\}\cap A'$
 
\noindent For, we eliminate $f_{12},f_3$ and $f_4$ from the set of equations ${\mathcal S}:=\{ Ec_1=0, Ec_2=0, U-(6f_{12}+3f_3+6f_4)=0,T-hessred=0 \}$, where $T$ and $U$ are auxiliary variables.

\begin{verbatim}

> Ec1 := (1/(-2*(-1 + x2)))*N1: 

> Ec2 := (1/(-2*x2*(-1 + x2)^2))*N2:

>hessred := -(N11*N22-N12*N21)/(4(x_2-1)^2):

>El:=eliminate({Ec1, Ec2, T-hessred, U + (-6*f12 - 6*f4 - 3*f3)}, {f12, f3, f4}):
 
>nops(El);
                               2
>El[1,1];

f12 = (96*U*x_2^4*x_3^2 + 24*U*x_2^3*x_3^3 + 24*U*x_2^4*x3 - 72*U*x_2^3*x_3^2 +

24*U*x_2^2*x_3^3 - 108*x_2^4*x_3^2 - 36*x_2^3*x_3^3 - 12*U*x_2^4 

- 54*U*x_2^3*x3 - 96*U*x_2^2*x_3^2 - 36*x_2^4*x3 + 72*x_2^3*x_3^2 

- 36*x_2^2*x_3^3 - 42*U*x_2^3 - 14*U*x_2^2*x3 - 8*U*x2*x_3^2 

+ 9*x_2^4 + 63*x_2^3*x3 + 108*x_2^2*x_3^2 + 14*U*x_2^2 + 

6*U*x2*x3 + 45*x_2^3 + 27*x_2^2*x3 + 82*U*x2

- 10*U*x3 - 9*x_2^2 + 9*x2*x3 + 38*U - 81*x2 + 9*x3 - 36)/

(3*(6*x_2^2*x3 + 3*x_2^2 + 6*x2*x3 + 1)*(2*x_2^2*x3 - 3*x_2^2 -2*x2*x3 - 4*x3 + 7))
 
>newf12:=eval(f12, El[1, 1]):

>newf3:=eval(f3,El[1,2]):

>newf4:=eval(f3,El[1,2]):
 
>degree(El[2, 1], T);
           1                    1
>beta1 := factor(coeff(El[2, 1], T, 1));
                      
          beta1 := beta1 := 9*(6*x_2^2*x3 + 3*x_2^2 + 6*x2*x3 + 1)*\\

(2*x_2^2*x3 - 3*x_2^2 - 2*x2*x3 - 4*x3 + 7)^2

>beta0 := factor(coeff(El[2, 1], T, 0));
\end{verbatim}

\noindent The output of the elimination gives us two lists  $[\rmEl[1,1],\rmEl[1,2],\rmEl[1,3]]$ and $[\rmEl[2,1]]$. The elements in the first list have the form $\rmEl[1,1]=f_{12}-(r_1/\delta)$, (see above), $\rmEl[1,2]= f_3-(r_2/\delta)$ 
and $\rmEl[1,3]=f_{4}-(r_3/\delta)$, where $r_i$ are  polynomials in $x_2,x_3,U$ of degree $1$ in $U$ and 
degrees $\leq 6$ in $x_2,x_3$, and $\delta$ is a polynomial in $x_2,x_3$ that is everywhere positive in ${\mathcal A}'$ (the code above).

\noindent The second list has only one element  $\rmEl[2,1]= \beta_1 T +\beta_0$, with $\beta_i$ polynomials in $x_2,x_3,U$.The polynomial $\beta_1$ dependes only of $x_2, x_3$ and is positive on ${\mathcal A}'$ (see above). Expanding  
$$\beta_0=  -4x_2^2(x_2-1)(2x_2x_3-x_2-1)(2x_2x_3+x_2+1)(2x_2^2x_3-3x_2^2+2x_2x_3-1)P$$
where $P=\alpha_2 U^2+\alpha_1 U+ \alpha_0$ is a polynomial of degree $2$ in $U$, and total degree  in $x_2,x_3$, $\leq 15 $ (see the Appendix). The factors of $\beta_0$,
 except for $P$, have a constant sign over ${\mathcal A}'$: four of them are negative and one of them is positive.

 \noindent We conclude that, for a given value of the parameters $(f_0,f_{12}, f_3,f_4)$, with $1=f_0+6f_{12},+3f_3+6f_4$ and a given critical point of $h_f$,  $(x_2,x_3)\in {\mathcal A}'$, by setting $T:=hessred(x_2,x_3)$ and $U:=1-f_0$, we get a solution of the system ${\mathcal S}$ and therefore, it holds $f_{12}=r_1/\delta, \ f_{3}=r_2/\delta, \ f_{4}=r_3/\delta$ and $\rmEl[2,1]=0$. 
  Reciprocally if $f_{12},f_{3},f_{4},U,x_2,x_3,T$ verify $f_{12}=r_1/\delta, \ f_{3}=r_2/\delta, \ f_{4}=r_3/\delta$ and $\rmEl[2,1]=0$, the point $(x_2,x_3)$ is a critical point of $h_f$ (with parameters, $f_0:=1-U, f_{12}, f_{3}, f_{4}$ and $T$ is the value of $hessred$ of $h_f$ at $(x_2,x_3)$.
  
\noindent As $\beta_1 >0$ over ${\mathcal A}\setminus \{ (1,1) \}$, the sign of $hessred$ at a critical point is the same as that of $-\beta_0$. Since the sign of $hessred$ is the opposite of the sign of the determinant of the Hessian of $h_f$,  in order to verify that the the determinant of the Hessian of $h_f$ is positive at every critical point of $h_f$ in ${\mathcal A}'$ (for parameters in the restricted admissible region), we have to prove that $\beta_0$ is positive at every critical point of $h_f$ in ${\mathcal A}'$ (with the values of the parameters in the restricted admissible region). Hence, we are led to  prove that the sign of $P = \alpha_2 (x_2,x_3)U^2 + \alpha_1(x_2,x_3)U + 
\alpha_0(x_2,x_3)$ is positive for $(x_2,x_3)\in {\mathcal A}'$ whenever $U\in [0,1]$, and  $ r_2\geq r_1 \geq r_3 \geq 0$, and $r_2 \geq r_1 +(2/3) r_3$ (inequalities that define the restricted admissible region).

\noindent Therefore we write:
$r_1-r_3=:c_1 U + c_0$, $r_3=: b_1 U+b_0$, and $r_2-(r_1+(2/3)r_3)=: a_1 U+a_0$,
where $c_j, b_j, a_j \in \QQ[x_2,x_3]$. Explicitly we find:

\[
 \begin{array}{lll}
c1&=&216x_2^4x_3^2+48x_2^3x_3^3+24x_2^4x_3-120x_2^3x_3^2+48x_2^2x_3^3-42x_2^4-\\
&&72x_2^3x_3-216x_2^2x_3^2-84x_2^3-24x_2^2x_3-40x_2x_3^2+40x_2^2-24x_2x_3+164x_2+82\\
c0&=&-252x_2^4x_3^2-72x_2^3x_3^3-36x_2^4x_3+108x_2^3x_3^2-72x_2^2x_3^3+45x_2^4+\\
&&108x_2^3x_3+252x_2^2x_3^2+90x_2^3+36x_2^2x_3+36x_2x_3^2-36x_2^2+36x_2x_3-162x_2-81\\
b1&=&-2(2x_2x_3-x_2-1)(30x_2^3x3+6x_2^2x_3^2+15x_2^3-9x_2^2x_3+6x_2x_3^2+6x_2^2\\
&&-24x_2x_3-19x_2-5x_3-22)\\
b0&=&9(2x_2x_3-x_2-1)(8x_2^3x_3+2x_2^2x_3^2+4x_2^3-x_2^2x_3+2x_2x_3^2+x_2^2-6x_2x_3\\
&&-4x_2-x_3-5)\\
a1&=&132x_2^4x_3^2-24x_2^3x_3^3-252x_2^4x3+264x_2^3x_3^2-24x_2^2x_3^3-159x_2^4+\\
&&270x_2^3x3-204x_2^2x_3^2+42x_2^3+142x_2^2x_3-256x_2x_3^2+136x_2^2-174x_2x_3-\\
&&82x_2+158x_3+31\\
a0&=&-180x_2^4x_3^2+36x_2^3x_3^3+324x_2^4x3-360x_2^3x_3^2+36x_2^2x_3^3+207x_2^4-\\
&&207x_2^3x3+180x_2^2x_3^2-45x_2^3-171x_2^2x_3+288x_2x_3^2-135x_2^2+135x_2x_3+81x_2\\
&&-153x_3-36\\
\end{array} \]

\noindent {\sc Second part of the proof}. We have a polynomial $P\in \QQ [x_2,x_3,U]$ of degree two in $U$ with coefficients of degree $\leq 7$ in $x_2,x_3$; $P= \alpha_2 U^2+ \alpha_1 U+ \alpha_0$ (see  the Appendix). 
   We consider the region ${\mathcal A}\times [0,1]$, and our goal is to prove the following :\\
   
\noindent {\bf CLAIM}:
   On the semi algebraic subset  of ${\mathcal A} \times [0,1] = \{ (x_2,x_3,u): 0\leq x_3\leq x_2 \leq 1, 0\leq U \leq 1 \}$ given by the inequalities below, the polynomial $P$ is non negative, and its zeroes are contained in $x_2=1$.
   
   \begin{align}
   a_1(x_2,x_3) \ U + a_0(x_2,x_3) \ \geq 0 \label{eq:1}\\
    b_1(x_2,x_3) \ U + b_0(x_2,x_3) \ \geq 0 \label{eq:2}\\
    c_1(x_2,x_3) \ U + c_0(x_2,x_3) \ \geq 0 \label{eq:3}
  \end{align}

This claim will be proved in some steps according the following lemma.
   
\begin{lemma}\label{primerlema}
\begin{itemize}
\item[]

\item[{\rm (1)}]\noindent  For $ (x_2,x_3) \in {\mathcal A}$ we have: \mbox{$c_1 \geq 0, c_0\leq 0$,  $b_1\leq 0, b_0\geq 0$}, and $c_0+c_1\geq 0$. Moreover $b_0$, $b_1$, $c_0$ and $c_1$, each of them vanishes only in ${\mathcal A}$ at the point $(1,1)$.
In ${\mathcal A}$ it holds that $a_1\leq 0 \Rightarrow a_0> 0$ and $ a_1(1, 0) <0 $.

\item[{\rm (2)}]\noindent ${\mathcal A} \cap \; \{-a_0 c_1+c_0 a_1 \leq 0, a_1 \geq 0 \} \subset \{  a_0 +a_1 \geq 0\}$.

\item[{\rm (3)}]\noindent  In ${\mathcal A} \; \cap \; \{a_1 \geq 0\} \; \cap \; 
\{-a_0 c_1+c_0 a_1 \geq 0\} \; \cap \;  \{a_0 +a_1 \geq 0\}
$, it holds ${\mathrm {max}}(0,-a0/a1,-c0/c1)=
 -a_0/a_1$. And  in  ${\mathcal A}\setminus \{(1,1)\} \; \cap \;  \{a_1 \geq 0\} \; \cap \; 
\{-a_0 c_1+c_0 a_1 \leq 0\}
$, it holds ${\mathrm {max}}(0,-a_0/a_1,-c_0/c_1)=-c_0/c_1$ 

\item[{\rm (4)}]\noindent  $X:={\mathcal A}\setminus \{(1,1)\} \; \cap \; \{a_1 \geq 0 \} \; \cap \;  \{a_0+a_1 \geq 0\}$  is contained in $\{b_0+b_1 \leq 0\}$, and hence on $X$ it holds ${\mathrm {min}}(1, -b0/b1)
=-b_0/b_1$. 

\end{itemize}

 \end{lemma}
 
 \vspace{2pt}
 
 \noindent {\bf Proof.}
 
 The following computations with the Maple package {\sc SolveTools} 
 prove all the  claims of the Lemma. For, (1) (first claim of (1)) and (2) come from ineqs1 to ineqs8, (3) comes from ineqs9 and (4) also from ineqs8. 
 
\begin{figure}[h!]

\centering

\includegraphics[width=0.6\textwidth]{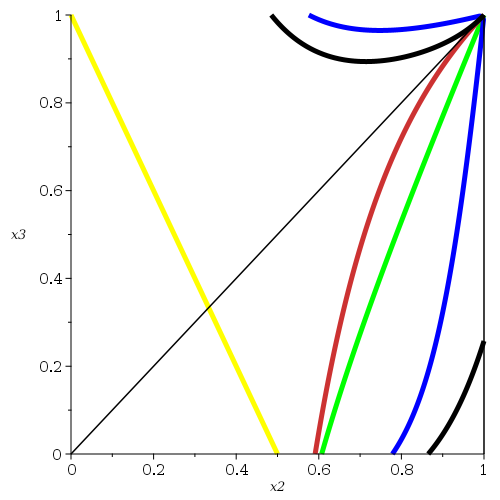}
\caption{$a_0=0$ blue(+), $a_1=0$ black(-), $a_0+a_1=0$ orange(+), $b_0+b_1=0$ yellow(-), $-a_0c_1+c_0a_1=0$ green(-).}
\label{fig:2d8}
\end{figure}

 \begin{verbatim}
 
> with(SolveTools);
  
>ineqs1:= {c1 = 0, 0 <= x3, x2 <= 1, x3 <= x2}:
                                                   
>sasolyx1:= SolveTools:-SemiAlgebraic(ineqs1, [x3, x2]);
               sasolyx1 := [[x3 = 1, x2 = 1]]
               
>ineqs2:= {c0 = 0, 0 <= x3, x2 <= 1, x3 <= x2}:
  
>sasolyx2:= SolveTools:-SemiAlgebraic(ineqs2, [x3, x2]);
                         sasolyx2 := [[x3 = 1, x2 = 1]]
                         
>ineqs3:= {0 <= x3, x2 <= 1, x3 <= x2, 0 < c0}:

>sasolyx3 := SolveTools:-SemiAlgebraic(ineqs3, [x3, x2]);
                         sasolyx3 := []
                         
>ineqs4 := {0 <= x3, x2 <= 1, x3 <= x2, c1 < 0}:

>sasolyx4 := SolveTools:-SemiAlgebraic(ineqs4, [x3, x2]);
                         sasolyx4 := []
    
>ineqs5 := {0 <= b1, 0 <= x3, x2 <= 1, x3 <= x2}:

>sasolyx5 := SolveTools:-SemiAlgebraic(ineqs5, [x3, x2]);
                         sasolyx5 := [[x3 = 1, x2 = 1]]

>ineqs6 := {0 <= x3, b0 <= 0, x2 <= 1, x3 <= x2}:

>sasolyx6 := SolveTools:-SemiAlgebraic(ineqs6, [x3, x2]);
                         sasolyx6 := [[x3 = 1, x2 = 1]]

> ineqs7 := {0 <= x3, x2 <= 1, x3 <= x2, c1 + c0 < 0};

>sasolyx7 := SolveTools:-SemiAlgebraic(ineqs7, [x3, x2]);
                         sasolyx7 := []

>ineqs8 := {0 <= x3, x2 <= 1, x3 <= x2, 0 <= a0 + a1, 0 < b0 + b1}:

>sasolyx8 := SolveTools:-SemiAlgebraic(ineqs8, [x3, x2]);
                         sasolyx8 := []
                         
>ineqs9 := {0 <= a1, 0 <= x3, x2 <= 1, x3 <= x2, -a0*c1 + a1*c0 <= 0, a0 + a1 < 0}:

>sasolyx9 := SolveTools:-SemiAlgebraic(ineqs9, [x3, x2]);
                         sasolyx9 := []
                         
>ineqs10 := {0 <= x3, 0 <= a0*b1 - a1*b0, a1 <= 0, x2 <= 1, x3 <= x2}:

>sasolyx10 := SolveTools:-SemiAlgebraic(ineqs10, [x3, x2]);
                sasolyx10 := [[x3 = 1, x2 = 1]]
                
>ineqs11 := {0 <= x3, a1 <= 0, x2 <= 1, x3 <= x2, -a0*c1 + a1*c0 >= 0}:

>sasolyx11 := SolveTools:-SemiAlgebraic(ineqs11, [x3, x2]);
                sasolyx11 := [[x3 = 1, x2 = 1]]
                
> ineqs12 := {0 <= x3, a1 <= 0, x2 <= 1, x3 <= x2, -a0*b1 + a1*b0 <= 0}:

>sasolyx12 := SolveTools:-SemiAlgebraic(ineqs12, [x3, x2]);
                sasolyx12 := [[x3 = 1, x2 = 1]]

\end{verbatim}
The  emptiness of these  semi algebraic sets above can be certified also by plotting the respective curves given by the equality to zero of the equations involved in the respective descriptions. See \autoref{fig:2d8}.
We write the  sign in parentheses to the right of the color is the sign of the polynomial that defines the colored curve in the region that contains the point $(x_2=1,x_3=0)$. Thus, in \autoref{fig:2d8} we see that in the trinagle${\mathcal A}$, the region  $a_1<0$ contains the point $(1,0)$ and it is contained in the region $a_0>0$; which proves the second claim of (i) in lemma \ref{primerlema}.

In the following we will use systematically the {\sc Plot} package of Maple to verify signs of polynomials in two variables
 in some subsets of our region ${\mathcal A}$. This package is a numerical package,  therefore more eficient than SolveTools (partially symbolic package) when working with polynomials of degree bigger than $2$. We can see in the Appendix the size of the $\alpha_i$'s we are dealt with.
 
  \begin{lemma} \label{segundolema}
  
  \begin{itemize}
  
\item[{\rm (i)}]\noindent  For $(x_2,x_3) \in {\mathcal A} \;  \cap \;  \{a_1\geq 0\} \; \cap \; \{a_0+a_1 <0 \}$, there is no $U\in [0,1]$ verifying \eqref{eq:1}. Moreover 
$ {\mathcal A} \;  \cap \;  \{ \; a_1\leq 0 \; \} \subset \{ \; a_0 + a_1\geq 0 \; \}$.

\item[{\rm (ii)}]\noindent  For $(x_2,x_3) \in {\mathcal A} \;  \cap \;  \{a_1\geq 0\} \; \cap \;  
\{-a_0 c_1+c_0 a_1 \geq 0\} \; \cap \; \{ a_0 +a_1 \geq 0 \}$ it holds that the  
inequalities  \eqref{eq:1}, \eqref{eq:2} and \eqref{eq:3} and $U\in [0,1]$ are equivalent to $U \in [-a_0/a_1, -b_0/b_1]$, that is 
$ -a_0/a_1 \leq U \leq -b_0/b_1$.

\item[{\rm (iii)}]\noindent For $(x_2,x_3) \in {\mathcal A} \; \cap \; \{ a_1\geq 0\} \; \cap 
\; \{ -a_0 c_1 + a_1 c_0 \leq 0 \} \cap \; \{ a_0 +a_1 \geq 0 \}$ it holds that 
the inequalities \eqref{eq:1}, \eqref{eq:2} and \eqref{eq:3} and $U\in [0,1]$ are equivalent to  $U \in [-c_0/c_1, -b_0/b_1]$.

\item[{\rm (iv)}]\noindent 
If $(x_2,x_3) \in {\mathcal A}\setminus\{(1,1)\} \; \cap \; \{ \; a_1\leq 0 \; \}$, it holds that the inequalities   \eqref{eq:1}, \eqref{eq:2}, \eqref{eq:3},  and $U \in [0,1]$ are equivalent to $U \in [-c_0/c_1,-b_0/b_1]$

\end{itemize}
\end{lemma}

 \noindent {\bf Proof.} 
 (i) Notice that for $(x_2, x_3, U)$ verifying \eqref{eq:1}, when $a_1\geq0$, we have  $U\geq -a_0/a_1$, and since $U\leq 1$, it must be $a_1+a_0\geq 0$, otherwise if $a_1+a_0 < 0$
  one gets $-a_0/a_1 > 1$, which is a contradiction. Hence for $(x_2, x_3, U)\in {\mathcal A}\times[0,1]$ verifying  \eqref{eq:1} and $a_1 \geq0$,  $(x_2,x_3)$ is to the right of the orange curve in \autoref{fig:2d8}. For the second assertion see \autoref{fig:2d8}.\\
  
 \noindent (ii) Now we observe  that $-c_0/c_1$ is positive in ${\mathcal A}\setminus\{(1,1)\}$, and $-b_0/b_1 \leq1$ holds to the right of the line yellow in particular to the right of the orange line in \autoref{fig:2d8}.
  
 \noindent For $(x_2, x_3, U)$ verifying \eqref{eq:2}, since $b_1<0$ in ${\mathcal A}\setminus \{(1,1)\}$, one gets 
  $U\leq-b_0/b_1$ and $U\leq 1$. So that $U\leq {\mathrm {min}}(1, -b_0/b_1)$. But,  $-b_0/b_1 \leq 1$ if and only if $b_0+b_1 \leq 0$ and this holds to the right of the curve yellow in \autoref{fig:2d8}.
 
 \noindent Finally, when $a_1 \geq 0$, that is to the left of the curve in black in \autoref{fig:2d8}, if $-a_0 c_1+c_0 a_1,  \geq 0$, left of the green curve (and right of the orange curve),  $-a_0/a_1 \geq -c_0/c_1$. So that for $(x_2, x_3, U)\in {\mathcal A}\setminus\{ (0,1)\}\times [0,1]$, with  $a_1(x_2,x_3) \geq 0$ and $ -a_0 c_1+c_0 a_1 \geq 0$,  $(x_2, x_3, U)$ verifies the \eqref{eq:1} if and only if $U\in [-a_0/a_1, -b_0/b_1]$.
 
 All these arguments  prove (ii) of the Lemma.\\
 
 \noindent Analogously we get (iii), when  $-a_0 c_1+c_0 a_1  \leq 0$. 
 
 \noindent For (iv), given $(x_2,x_3) \in {\mathcal A}' \; \cap \; \{a_1\leq 0\}$, and $(x_2,x_3,U)$ verifying \eqref{eq:1}, \eqref{eq:2}, \eqref{eq:3}, since $b_1 \leq 0 $ and $c_1 \geq 0$, $U \geq -c_0/c_1 \geq 0$ and $U \leq -b_0/b_1\leq 1$. 
 Reciprocally 
 if $(x_2,x_3) \in {\mathcal A}'$ and $-c_0/c_1 \leq U \leq  \-b_0/b_1$, we have $U\in [0,1]$, because $-c0/c1 \geq 0$ and $-b0/b1\leq 1$ (since $b_0+b_1 \leq 0$ and $b_1< 0$). Moreover since $c_1>0$ and $b_1<0$ we have that \eqref{eq:2} and \eqref{eq:3} hold.
 Finally, looking at ``ineqs12'', we have that $-a_0 b_1+a_1b_0 \geq 0$, hence $-b_0/b_1\leq -a_0/a_1$ and, since $a_1<0$, one gets that \eqref{eq:1} holds.

 \begin{lemma}\label{tercerlema}
 
 Let be $P= \alpha_2 U^2+\alpha_1 U + \alpha_0 \in \RR[U]$, with $\alpha_2\neq 0$, 
 and $\delta< \eta\in \RR$ and $I:=[\delta, \eta]$. Then  
 
\begin{itemize}
\item\noindent  If $\alpha_2<0$, $P$ is positive everywhere in the interval $I$ if and only if $P(\delta)>0, P(\eta) >0 $. 
 \item\noindent  If $\alpha_2>0$, $P$ is positive everywhere in the interval $I$, if and only if $P(\delta)>0$, $ P(\eta) >0 $, and ${\mathrm {sign}}(P'(\delta))={\mathrm {sign}}(P'(\eta))$.
\end{itemize}
\end{lemma}
\noindent {\bf Proof.} High school exercise.

\begin{lemma}\label{cuartolema}

 With the same notations:
 
\noindent {\rm (i)} $P(x_2,x_3,-a0/a1)$ is non negative for $(x_2,x_3)\in {\mathcal A}' \ \cap  
\{-a_0 c_1 + c_0 a_1 \geq 0 \} \cap \{ a_0+a_1 \geq 0 \}$ .\\

\noindent {\rm (ii)} $P(x_2,x_3,-c_0/c_1)$ is non negative for $(x_2,x_3)\in {\mathcal A}'
  $.\\

\noindent {\rm (iii)}  $P(x_2,x_3,-b_0/b_1)$ is non negative for $(x_2,x_3)\in {\mathcal A}' \cap a_0+a_1 > 0  \ $.\\

\noindent {\rm (iv)} $P'(x_2,x_3,-b_0/b_1)<0$ for $(x_2,x_3)\in {\mathcal A}' \cap \{ \; a_0+a_1>0 \; \}\cap  
\{ \; a_1 \geq 0 \;\} \cap \{ \; \alpha_2 \geq 0 \; \}.$\\

\noindent {\rm (v)} $P'(x_2,x_3,-a_0/a_1)<0$  for $(x_2,x_3)\in {\mathcal A}'\cap \{ \; a_1(x_2,x_3)>0 \; \} \cap  
\{ \; -a_0 c_1+c_0 a_1 \geq 0 \;\} \cap \{ \; \alpha_2 \geq 0 \; \}.$\\
 
\noindent {\rm (vi)} $P'(x_2,x_3,-c_0/c_1)<0$  for $(x_2,x_3)\in {\mathcal A}'\cap \{ \;a_1(x_2,x_3)>0 \; \}\cap  
\{ \; -a_0c_1+c_0a_1 \leq 0 \; \}$.\\

\end{lemma}

\noindent {\bf {Proof. of the Lemma}}

\begin{figure}[h!]

\centering

\includegraphics[width=0.5\textwidth]{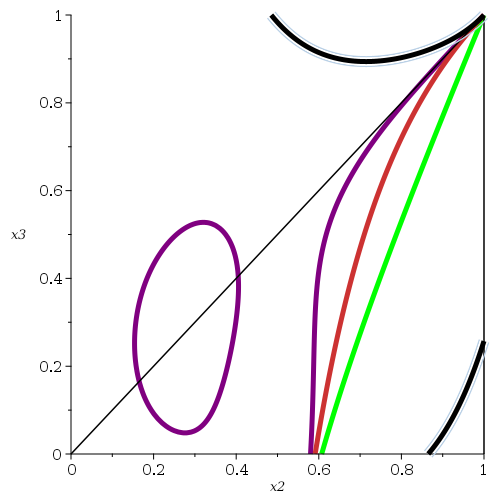}
\caption{ $a_0^2 \alpha_2 - a_0 a_1\alpha_1 + a_1^2 \alpha_0=0$ purple(+), $a_1=0$ black(-), $a_0+a_1=0$ orange(+), $-a_0c_1+c_0a_1=0$ green(-).}
\label{fig:2dBa}
\end{figure}

\noindent (i)
 The sign of $P(x2,x3, -a_0/a_1)$ is the sign of $a_0^2 \alpha_2 - a_0 a_1 \alpha_1 + a_1^2 \alpha_0$. We  use ${\sc Plot}$ of Maple and see it in \autoref{fig:2dBa}.
 
 \begin{verbatim}
 
 >subs(x2 = 1, x3 = 0, a0^2*alpha2 - a0*a1*alpha1 + a1^2*alpha0);
 
                            74317824

> implicitplot([a0^2*alpha2 - a0*a1*alpha1 + a1^2*alpha0, -a0*c1 + a1*c0,

a0 + a1, x2 - x3, x2 - 1, a1], x2 = 0 .. 1, x3 = 0 .. 1, color = [purple, green,

orange, black, black, black]);
  
 \end{verbatim}

\begin{figure}[h!]

\centering

\includegraphics[width=0.5\textwidth]{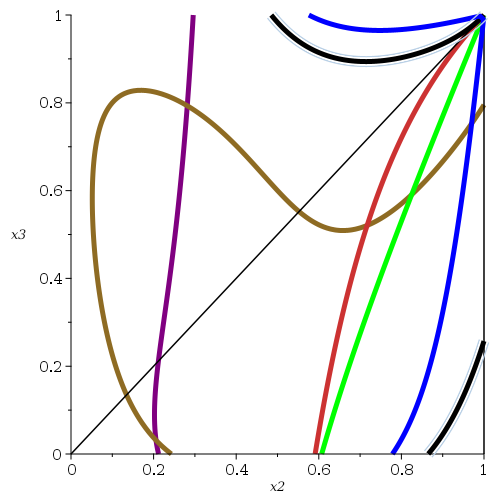}

\caption{ $b_0^2 \ \alpha_2 - b_0 \ b_1 \ \alpha_1 + b_1^2 \ \alpha_0 =0$ purple(+), $a_0 + a_1=0$ orange(+), $a_1 \ = \ 0$
black(-),  $a_0 \ =  0$ blue (+), $\alpha_2 \ =   0$ ``sienna''brown (+).}
\label{fig:2dBb}
\end{figure}

\noindent (iii)
  With the same argument as before, see \autoref{fig:2dBb}.
  
\begin{verbatim}
   
>subs(x2 = 1, x3 = 0, alpha0*b1^2 - alpha1*b0*b1 + alpha2*b0^2);

                            42467328

>implicitplot([alpha0*b1^2 - alpha1*b0*b1 + alpha2*b0^2, a0 + a1, alpha2, a1,

x2 - x3, x2 - 1], x2 = 0..1, x3 = 0..1, color = [purple, orange, sienna, black,

black, black]);
\end{verbatim}

\begin{figure}[ht]
 
\centering
 
\includegraphics[width=0.5\textwidth]{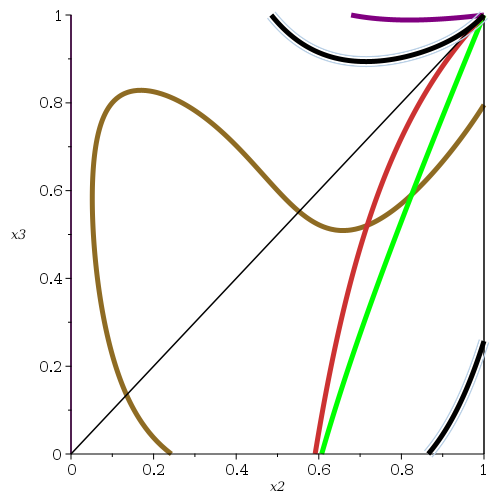}

\caption{ $c_0^2 \ \alpha_2 - c_0 \ c_1 \ \alpha_1 + c_1^2 \ \alpha_0=0$ purple(+), 
$a_0+a_1=0$ orange(+),$a_1 \ = 0$ black(-), $-a_0 \ c_1 \ + \ c_0 \ a_1 \ =0$ green(+).}
\label{fig:2dBc}
\end{figure}

 \noindent (ii) See \autoref{fig:2dBc}

 \begin{verbatim}
 
> subs(x2 = 1, x3 = 0, alpha0*c1^2 - alpha1*c0*c1 + alpha2*c0^2);

                           169869312
                           
>implicitplot([alpha0*c1^2 - alpha1*c0*c1 + alpha2*c0^2, alpha2, 
  
  -a0*c1 + a1*c0, a0 + a1, x2 - x3, x2 - 1, a1], x2 = 0..1, x3 = 0..1, 
  
  color = [purple, sienna, green, orange, black, black, black]);
  
>subs(x2 = 1, x3 = 0, alpha0*c1^2 - alpha1*c0*c1 + alpha2*c0^2);
  
 \end{verbatim}
 
\begin{figure}[ht]

\centering
 
\includegraphics[width=0.5\textwidth]{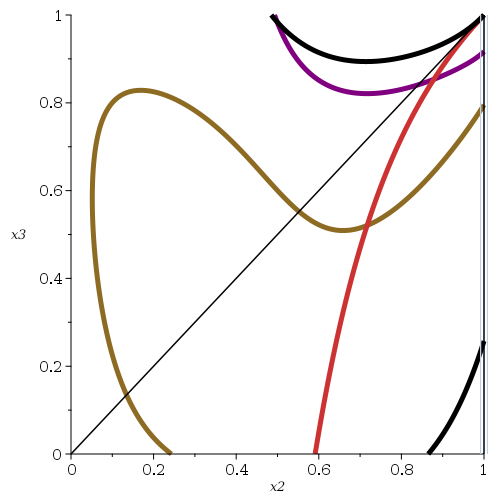}

\caption{ $\alpha_1 b_1 - 2 \alpha_2 b_0 \ =0$ purple(+), $a_0+a_1=0$ orange(+),$a_1 \ = 0$ black(-), $\alpha_2=0$ brown ``sienna'' (+).} \label{fig:2dBderb}
\end{figure}

\noindent (iv) Since $b_1$ is negative in ${\mathcal A}'$, the sign of the derivative 
of $P$ in $U=-b_0/b_1$ is the contrary of the sign of $\alpha_1 b_1 - 2 \alpha_2 b_0$. 
See below \autoref{fig:2dBderb}.

\begin{verbatim}
 
 > subs(x2 = 1, x3 = 0, alpha1*b1 - 2*alpha2*b0);
                  
                  4718592
                            
>implicitplot([alpha1*b1 - 2*alpha2*b0, alpha2,  a0 + a1,  x2 - x3, x2 - 1, a1], 

x2 = 0 .. 1, x3 = 0 .. 1, color = [purple,sienna,  orange, black, black, black]);

>subs(x2=0.8,x3=0,alpha1*b1-2*alpha2*b0):

\end{verbatim}

\begin{figure}[h!]

\centering
 
\includegraphics[width=0.5\textwidth]{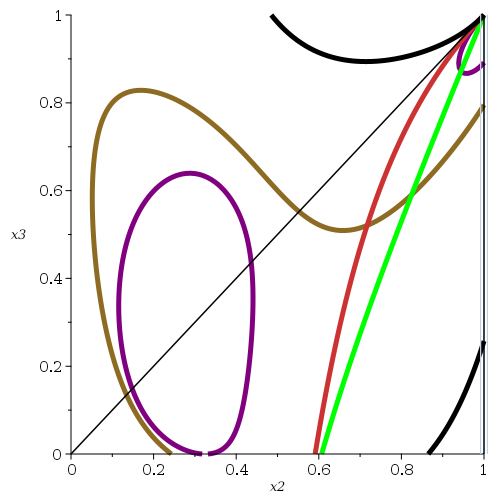}

\caption{ $\alpha_1 a_1 - 2 \alpha_2 a_0 \ =0$ purple(-), $a_0+a_1=0$  orange(+), $-a_0 c_1 + a_1 c_0=0$ green(+), $a_1 \ =\ 0$ black(-), $\alpha_2=0$ brown ``sienna'' (+).}
\label{fig:2dBdera}
\end{figure}

\noindent (v) See \autoref{fig:2dBdera}

\begin{verbatim}
 > subs(x2 = 1, x3 = 0, {alpha2, -2*a0*alpha2 + a1*alpha1});
 
                       {-5013504, 79872}

> implicitplot([-2*a0*alpha2 + a1*alpha1, alpha2, -a0*c1 + a1*c0, a0 + a1, \\

x2 - x3, x2 - 1, a1], x2 = 0 .. 1, x3 = 0 .. 1, 

color = [purple, sienna, green, orange, black, black, black]);
\end{verbatim}

\begin{figure}[h!]

\centering
 
\includegraphics[width=0.5\textwidth]{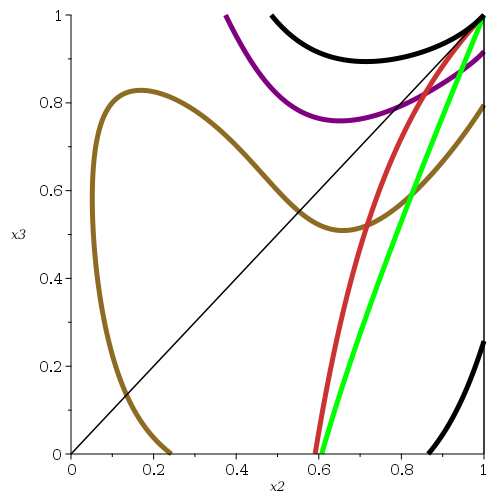}

\caption{ $\alpha_1 c_1 - 2 \alpha_2 c_0 \ =0$ purple(-), $a_0+a_1=0$ orange(+),$a_1 \ = 0$ black(-), 
$-a_0 c_1 + a_1 c_0=0$ green(+), $\alpha_2=0$ brown ``sienna'' (+).}
\label{fig:2dBderc}
\end{figure}

\newpage
\noindent (vi) See \autoref{fig:2dBderc}
\begin{verbatim}
 > subs(x2 = 1, x3 = 0, {alpha2, alpha1*c1 - 2*alpha2*c0});
 
                       {-9437184, 79872}

> implicitplot([alpha1*c1 - 2*alpha2*c0, alpha2, a0 + a1, -a0*c1 + a1*c0, x2 -x3,

x2 - 1, a1], x2 = 0 .. 1, x3 = 0 .. 1, color = [purple, sienna, orange, green,

black, black, black]);

\end{verbatim}

\begin{proposition}
 Let $\Sigma$ be  the following semi algebraic subset of ${\mathcal A}'\times [0,1]$, 
\begin{align*}
 \Sigma:=\{(x_2,x_3,U) \in
A' \times [0,1]: \;  a_1.U+a_0 \geq 0, \; b_1(x_2,x_3) U+b_0(x_2,x_3)\geq 0, \\
c_1(x_2,x_3) U+c_0(x_2,x_3) \geq 0\}
\end{align*}

Then, $P$ is positive in $\Sigma$.
\end{proposition}

\noindent{\bf Proof. of the proposition. } \ \
Let us set  $\Sigma_{a_1}:=  \Sigma \cap a_1 >0$ and 
 $\Sigma_{-a_1}:=  \Sigma \cap a_1 <0$ . 

\noindent We examine the value of $P$ in $(x_2,x_3,U)\in \Sigma_{a_1}$. 

Assume first that $\alpha_2(x_2,x_3)>0$.
Since $a_1>0$ , and $1\geq U\geq -a_0/a_1$, we have that $a_0+a_1>0$. 
If 
$-a_0 c_1+c_0 a_1 \geq 0$, then by \ref{segundolema}, $U \in [-a_0/a_1, -b_0/b_1]$. In that case, by \ref{tercerlema}  and \ref{cuartolema}, (iii), (iv) and (v) we get  that $P$ is positive in $(x_2,x_3,U)$.
Otherwise, if  
$-a_0 c_1+c_0 a_1 \leq 0$, then $U \in [-c_0/c_1, -b_0/b_1]$, (again we have $a_0+a_1\geq 0$), and by \ref{tercerlema} and \ref{cuartolema} (ii), (iii), (iv) and (vi)  proof  that $P$ is positive in $(x_2,x_3,U)$.

Assume now $\alpha_2 (x_2,x_3) <0$, then in  case $-a_0 c_1+c_0 a_1 \geq 0$ (resp.  in case $-a_0 c_1+c_0 a_1 \leq 0$), claims  (i) and (ii) (resp. (i) and (iii)) assure that $P$ is positive in $(x_2,x_3,U)$.

\noindent We examine the case   $(x_2,x_3,U) \in \Sigma_{-a_1}$. 

In that case   
$\alpha_2(x_2,x_3) >0$ see \autoref{fig:2dBderc}. By \ref{segundolema} (i) we know $a_0+a_1 \geq 0$, also by \autoref{fig:2dBderc}  it holds that $-a_0c_1+a_1c_0 >0$. Finally  by \ref{segundolema} (iv)  $U\in [-c_0/c_1,-b_0/b_1]$. Therefore we can use  \ref{tercerlema}, and  (ii), (iii), (iv), (vi) \ref{cuartolema} to prove that $P(x_2,x_3,U)$ is positive.   

Summing up, using the last two lemmas we get the Proposition.

\medskip

\begin{remark}\label{notabiologica}

{\rm We can improve our result, proving that if  ${\mathcal C}$ is the curve  of equation $g(x_2,x_3):=  {\sf a_3}(x_2,x_3)- {\sf a_2}(x_2,x_3)- (2/3) {\sf a_4}(x_2,x_3)$, that is, $g=-2x_2^2x_3 + 7x_2^2 - 4x_2x_3 -1$, then the maximum of $h_f$ in ${\mathcal A}'$ belongs to the subset 
${\mathcal A}' \cap \{  g(x_2,x_3) \geq 0 \}$ (for realistic values of $f_0$, as $f_0 \geq 0 .7$). Indeed, using all the work done one has just to verify that the gradient of $h_f$ along this curve points towards the point $(x_2=1,x_3=0).$
The biological sense of this inequality seems to be sense since in the three examples of realistic applications at the end of the paper \cite{ChorHendySnir2006}: 
 ((rat, mouse), Human), ((guinea pig, goose), C. elegans) and ((D. melangoster, H. vulgaris), Human) the maximum obtained holds this inequality.} \footnote{The authors provide only the frecuencies of the  observed data in one of the cases. In the other two cases they provide just the maximum, claiming that they are unique. In the first case we have used the frecuencies given in the cited paper, and in the two other cases, we have used the probabilities provided by the model, evaluated at  the given maximum $(x_2^{*},x_3^{*})$;  we have verified in all the three cases that the inequelity $g(x_2^{*},x_3^{*}) \geq 0$ holds.}

\end{remark}


\section*{Appendix}

Here there are some of the data to appreciate the size of the computations.

\medskip

\noindent $alpha2 := -3456*x2^9*x3^6 + 17280*x2^{10}*x3^4 - 20736*x2^9*x3^5 - 17280*x2^8*x3^6 + 77760*x2^{10}*x3^3 + 152064*x2^9*x3^4 + 10368*x2^8*x3^5 - 34560*x2^7*x3^6 + 64800*x2^{10}*x3^2 - 32832*x2^9*x3^3 - 101952*x2^8*x3^4 + 89856*x2^7*x3^5 - 34560*x2^6*x3^6 + 15120*x2^{10}*x3 - 137592*x2^9*x3^2 - 298080*x2^8*x3^3 - 314496*x2^7*x3^4 + 34560*x2^6*x3^5 - 17280*x2^5*x3^6 - 25920*x2^9*x3 - 85320*x2^8*x3^2 + 122688*x2^7*x3^3 + 212544*x2^6*x3^4 - 20736*x2^5*x3^5 - 3456*x2^4*x3^6 + 7236*x2^9 - 10368*x2^8*x3 + 398160*x2^7*x3^2 + 295776*x2^6*x3^3 + 202752*x2^5*x3^4 + 38016*x2^4*x3^5 - 5076*x2^8 - 21888*x2^7*x3 - 79920*x2^6*x3^2 - 124416*x2^5*x3^3 - 181440*x2^4*x3^4 + 34560*x2^3*x3^5 - 94176*x2^7 + 66816*x2^6*x3 - 575136*x2^5*x3^2 - 36000*x2^4*x3^3 - 86400*x2^3*x3^4 + 6912*x2^6 + 74304*x2^5*x3 + 264960*x2^4*x3^2 + 124736*x2^3*x3^3 + 7488*x2^2*x3^4 + 200520*x2^5 - 80992*x2^4*x3 + 238128*x2^3*x3^2 + 56800*x2^2*x3^3 + 56376*x2^4 - 193792*x2^3*x3 - 121584*x2^2*x3^2 + 6080*x2*x3^3 - 116480*x2^3 + 8304*x2^2*x3 - 95592*x2*x3^2 - 2336*x2^2 + 57728*x2*x3 - 10168*x3^2 + 33620*x2 + 14432*x3 - 6724$\\

\noindent $alpha1 := 10368*x2^9*x3^6 - 46656*x2^{10}*x3^4 + 51840*x2^9*x3^5 + 51840*x2^8*x3^6 - 184032*x2^{10}*x3^3 - 383616*x2^9*x3^4 - 41472*x2^8*x3^5 + 103680*x2^7*x3^6 - 149040*x2^{10}*x3^2 + 47520*x2^9*x3^3 + 164160*x2^8*x3^4 - 273024*x2^7*x3^5 + 103680*x2^6*x3^6 - 34344*x2^{10}*x3 + 325512*x2^9*x3^2 + 755136*x2^8*x3^3 + 787968*x2^7*x3^4 - 138240*x2^6*x3^5 + 51840*x2^5*x3^6 + 78624*x2^9*x3 + 210600*x2^8*x3^2 - 176256*x2^7*x3^3 - 333504*x2^6*x3^4 + 31104*x2^5*x3^5 + 10368*x2^4*x3^6 - 10692*x2^9 + 9288*x2^8*x3 - 954576*x2^7*x3^2 - 747648*x2^6*x3^3 - 487296*x2^5*x3^4 - 96768*x2^4*x3^5 + 11988*x2^8 + 35712*x2^7*x3 + 134928*x2^6*x3^2 + 189504*x2^5*x3^3 + 368064*x2^4*x3^4 - 86400*x2^3*x3^5 + 200448*x2^7 - 118800*x2^6*x3 + 1339200*x2^5*x3^2 + 58176*x2^4*x3^3 + 221184*x2^3*x3^4 - 30240*x2^6 - 179136*x2^5*x3 - 402624*x2^4*x3^2 - 289152*x2^3*x3^3 - 13824*x2^2*x3^4 - 423720*x2^5 + 110736*x2^4*x3 - 473904*x2^3*x3^2 - 121248*x2^2*x3^3 - 125784*x2^4 + 402048*x2^3*x3 + 241056*x2^2*x3^2 - 11232*x2*x3^3 + 226656*x2^3 - 11592*x2^2*x3 + 187704*x2*x3^2 + 1728*x2^2 - 116064*x2*x3 + 20376*x3^2 - 66420*x2 - 29016*x3 + 13284$\\

\noindent $alpha0 := -7776*x2^9*x3^6 + 31104*x2^{10}*x3^4 - 31104*x2^9*x3^5 - 38880*x2^8*x3^6 + 108864*x2^{10}*x3^3 + 241056*x2^9*x3^4 + 38880*x2^8*x3^5 - 77760*x2^7*x3^6 + 85536*x2^{10}*x3^2 - 10368*x2^9*x3^3 - 50544*x2^8*x3^4 + 207360*x2^7*x3^5 - 77760*x2^6*x3^6 + 19440*x2^{10}*x3 - 193590*x2^9*x3^2 - 472392*x2^8*x3^3 - 487296*x2^7*x3^4 + 129600*x2^6*x3^5 - 38880*x2^5*x3^6 - 57024*x2^9*x3 - 137538*x2^8*x3^2 + 29808*x2^7*x3^3 + 92016*x2^6*x3^4 - 7776*x2^4*x3^6 + 2673*x2^9 + 2268*x2^8*x3 + 566676*x2^7*x3^2 + 454248*x2^6*x3^3 + 277344*x2^5*x3^4 + 59616*x2^4*x3^5 - 5589*x2^8 - 7776*x2^7*x3 - 43740*x2^6*x3^2 - 64800*x2^5*x3^3 - 182736*x2^4*x3^4 + 51840*x2^3*x3^5 - 106920*x2^7 + 52488*x2^6*x3 - 763992*x2^5*x3^2 - 30456*x2^4*x3^3 - 134784*x2^3*x3^4 + 24624*x2^6 + 110160*x2^5*x3 + 142560*x2^4*x3^2 + 164592*x2^3*x3^3 + 6480*x2^2*x3^4 + 223074*x2^5 - 29808*x2^4*x3 + 234252*x2^3*x3^2 + 64152*x2^2*x3^3 + 69822*x2^4 - 207360*x2^3*x3 - 119556*x2^2*x3^2 + 5184*x2*x3^3 - 110160*x2^3 + 3240*x2^2*x3 - 92178*x2*x3^2 + 648*x2^2 + 58320*x2*x3 - 10206*x3^2 + 32805*x2 + 14580*x3 - 6561$\\

\noindent $N1 := -2*(-1 + x2)*(240*f12*x2^6*x3^4 + 96*f3*x2^6*x3^4 + 240*f4*x2^6*x3^4 - 240*f12*x2^6*x3^3 + 72*f12*x2^5*x3^4 - 96*f3*x2^6*x3^3 + 48*f3*x2^5*x3^4 - 240*f4*x2^6*x3^3 + 72*f4*x2^5*x3^4 - 48*x2^6*x3^4 - 240*f12*x2^6*x3^2 + 252*f12*x2^5*x3^3 - 288*f12*x2^4*x3^4 - 96*f3*x2^6*x3^2 + 120*f3*x2^5*x3^3 - 96*f3*x2^4*x3^4 - 240*f4*x2^6*x3^2 + 228*f4*x2^5*x3^3 - 288*f4*x2^4*x3^4 + 48*x2^6*x3^3 - 24*x2^5*x3^4 + 60*f12*x2^6*x3 + 42*f12*x2^5*x3^2 - 68*f12*x2^4*x3^3 - 120*f12*x2^3*x3^4 + 24*f3*x2^6*x3 + 12*f3*x2^5*x3^2 - 104*f3*x2^4*x3^3 - 48*f3*x2^3*x3^4 + 60*f4*x2^6*x3 + 66*f4*x2^5*x3^2 - 92*f4*x2^4*x3^3 - 120*f4*x2^3*x3^4 + 48*x2^6*x3^2 - 36*x2^5*x3^3 + 48*x2^4*x3^4 + 45*f12*x2^6 + 57*f12*x2^5*x3 - 84*f12*x2^4*x3^2 + 68*f12*x2^3*x3^3 + 18*f3*x2^6 + 18*f3*x2^5*x3 - 48*f3*x2^4*x3^2 + 40*f3*x2^3*x3^3 + 45*f4*x2^6 + 75*f4*x2^5*x3 - 84*f4*x2^4*x3^2 + 92*f4*x2^3*x3^3 - 12*x2^6*x3 - 6*x2^5*x3^2 + 12*x2^4*x3^3 + 24*x2^3*x3^4 + 45*f12*x2^5 - 25*f12*x2^4*x3 + 224*f12*x2^3*x3^2 + 52*f12*x2^2*x3^3 + 18*f3*x2^5 - 10*f3*x2^4*x3 + 72*f3*x2^3*x3^2 + 40*f3*x2^2*x3^3 + 45*f4*x2^5 - 43*f4*x2^4*x3 + 208*f4*x2^3*x3^2 + 76*f4*x2^2*x3^3 - 9*x2^6 - 15*x2^5*x3 + 24*x2^4*x3^2 - 12*x2^3*x3^3 - 42*f12*x2^4 - 50*f12*x2^3*x3 + 148*f12*x2^2*x3^2 - 12*f3*x2^4 + 12*f3*x2^3*x3 + 48*f3*x2^2*x3^2 - 42*f4*x2^4 - 38*f4*x2^3*x3 + 148*f4*x2^2*x3^2 - 9*x2^5 + 3*x2^4*x3 - 36*x2^3*x3^2 - 12*x2^2*x3^3 - 42*f12*x2^3 - 34*f12*x2^2*x3 + 38*f12*x2*x3^2 - 12*f3*x2^3 - 4*f3*x2^2*x3 + 12*f3*x2*x3^2 - 42*f4*x2^3 - 46*f4*x2^2*x3 + 30*f4*x2*x3^2 + 6*x2^4 + 6*x2^3*x3 - 24*x2^2*x3^2 - 19*f12*x2^2 - 55*f12*x2*x3 - 6*f3*x2^2 - 30*f3*x2*x3 - 19*f4*x2^2 - 53*f4*x2*x3 + 6*x2^3 + 6*x2^2*x3 - 6*x2*x3^2 - 19*f12*x2 - 17*f12*x3 - 6*f3*x2 - 10*f3*x3 - 19*f4*x2 - 19*f4*x3 + 3*x2^2 + 9*x2*x3 + 3*x2 + 3*x3)$\\

\noindent $N2 := -2*x2*(-1 + x2)^2*(120*f12*x2^5*x3^3 + 48*f3*x2^5*x3^3 + 120*f4*x2^5*x3^3 - 180*f12*x2^5*x3^2 + 240*f12*x2^4*x3^3 - 120*f3*x2^5*x3^2 + 96*f3*x2^4*x3^3 - 204*f4*x2^5*x3^2 + 240*f4*x2^4*x3^3 - 24*x2^5*x3^3 - 30*f12*x2^5*x3 - 156*f12*x2^4*x3^2 + 120*f12*x2^3*x3^3 - 12*f3*x2^5*x3 - 120*f3*x2^4*x3^2 + 48*f3*x2^3*x3^3 - 6*f4*x2^5*x3 - 228*f4*x2^4*x3^2 + 120*f4*x2^3*x3^3 + 36*x2^5*x3^2 - 48*x2^4*x3^3 + 45*f12*x2^5 - 168*f12*x2^4*x3 - 28*f12*x2^3*x3^2 + 30*f3*x2^5 - 48*f3*x2^4*x3 - 40*f3*x2^3*x3^2 + 63*f4*x2^5 - 120*f4*x2^4*x3 - 100*f4*x2^3*x3^2 + 6*x2^5*x3 + 36*x2^4*x3^2 - 24*x2^3*x3^3 + 153*f12*x2^4 - 220*f12*x2^3*x3 - 52*f12*x2^2*x3^2 + 90*f3*x2^4 - 72*f3*x2^3*x3 - 40*f3*x2^2*x3^2 + 171*f4*x2^4 - 188*f4*x2^3*x3 - 76*f4*x2^2*x3^2 - 9*x2^5 + 24*x2^4*x3 + 12*x2^3*x3^2 + 174*f12*x2^3 - 152*f12*x2^2*x3 + 100*f3*x2^3 - 48*f3*x2^2*x3 + 186*f4*x2^3 - 136*f4*x2^2*x3 - 27*x2^4 + 36*x2^3*x3 + 12*x2^2*x3^2 + 102*f12*x2^2 - 38*f12*x2*x3 + 60*f3*x2^2 - 12*f3*x2*x3 + 114*f4*x2^2 - 30*f4*x2*x3 - 30*x2^3 + 24*x2^2*x3 + 53*f12*x2 + 30*f3*x2 + 55*f4*x2 - 18*x2^2 + 6*x2*x3 + 17*f12 + 10*f3 + 19*f4 - 9*x2 - 3)$\\

\noindent $N11:=-3360*f12*x2^6*x3^4 - 1344*f3*x2^6*x3^4 - 3360*f4*x2^6*x3^4 + 3360*f12*x2^6*x3^3 + 2016*f12*x2^5*x3^4 + 1344*f3*x2^6*x3^3 + 576*f3*x2^5*x3^4 + 3360*f4*x2^6*x3^3 + 2016*f4*x2^5*x3^4 + 672*x2^6*x3^4 + 3360*f12*x2^6*x3^2 - 5904*f12*x2^5*x3^3 + 3600*f12*x2^4*x3^4 + 1344*f3*x2^6*x3^2 - 2592*f3*x2^5*x3^3 + 1440*f3*x2^4*x3^4 + 3360*f4*x2^6*x3^2 - 5616*f4*x2^5*x3^3 + 3600*f4*x2^4*x3^4 - 672*x2^6*x3^3 - 288*x2^5*x3^4 - 840*f12*x2^6*x3 - 3384*f12*x2^5*x3^2 + 3200*f12*x2^4*x3^3 - 1344*f12*x2^3*x3^4 - 336*f3*x2^6*x3 - 1296*f3*x2^5*x3^2 + 2240*f3*x2^4*x3^3 - 384*f3*x2^3*x3^4 - 840*f4*x2^6*x3 - 3672*f4*x2^5*x3^2 + 3200*f4*x2^4*x3^3 - 1344*f4*x2^3*x3^4 - 672*x2^6*x3^2 + 1008*x2^5*x3^3 - 720*x2^4*x3^4 - 630*f12*x2^6 + 36*f12*x2^5*x3 + 1260*f12*x2^4*x3^2 - 1088*f12*x2^3*x3^3 - 720*f12*x2^2*x3^4 - 252*f3*x2^6 + 72*f3*x2^5*x3 + 600*f3*x2^4*x3^2 - 1152*f3*x2^3*x3^3 - 288*f3*x2^2*x3^4 - 630*f4*x2^6 - 180*f4*x2^5*x3 + 1500*f4*x2^4*x3^2 - 1472*f4*x2^3*x3^3 - 720*f4*x2^2*x3^4 + 168*x2^6*x3 + 648*x2^5*x3^2 - 480*x2^4*x3^3 + 192*x2^3*x3^4 + 820*f12*x2^4*x3 - 2464*f12*x2^3*x3^2 + 96*f12*x2^2*x3^3 + 280*f3*x2^4*x3 - 960*f3*x2^3*x3^2 + 1180*f4*x2^4*x3 - 2336*f4*x2^3*x3^2 + 96*f4*x2^2*x3^3 + 126*x2^6 + 36*x2^5*x3 - 300*x2^4*x3^2 + 192*x2^3*x3^3 + 144*x2^2*x3^4 + 870*f12*x2^4 + 200*f12*x2^3*x3 + 456*f12*x2^2*x3^2 + 208*f12*x2*x3^3 + 300*f3*x2^4 - 176*f3*x2^3*x3 + 144*f3*x2^2*x3^2 + 160*f3*x2*x3^3 + 870*f4*x2^4 - 40*f4*x2^3*x3 + 360*f4*x2^2*x3^2 + 304*f4*x2*x3^3 - 180*x2^4*x3 + 480*x2^3*x3^2 - 96*f12*x2^2*x3 + 440*f12*x2*x3^2 + 96*f3*x2^2*x3 + 144*f3*x2*x3^2 + 48*f4*x2^2*x3 + 472*f4*x2*x3^2 - 150*x2^4 - 24*x2^3*x3 - 72*x2^2*x3^2 - 48*x2*x3^3 - 138*f12*x2^2 + 84*f12*x2*x3 + 76*f12*x3^2 - 36*f3*x2^2 + 104*f3*x2*x3 + 24*f3*x3^2 - 138*f4*x2^2 + 28*f4*x2*x3 + 60*f4*x3^2 - 72*x2*x3^2 - 76*f12*x3 - 40*f3*x3 - 68*f4*x3 + 18*x2^2 - 12*x2*x3 - 12*x3^2 - 38*f12 - 12*f3 - 38*f4 + 12*x3+6 $\\

\noindent $N12=N21 = -2*(x2 - 1)*(960*f12*x2^6*x3^3 + 384*f3*x2^6*x3^3 + 960*f4*x2^6*x3^3 - 720*f12*x2^6*x3^2 + 288*f12*x2^5*x3^3 - 288*f3*x2^6*x3^2 + 192*f3*x2^5*x3^3 - 720*f4*x2^6*x3^2 + 288*f4*x2^5*x3^3 - 192*x2^6*x3^3 - 480*f12*x2^6*x3 + 756*f12*x2^5*x3^2 - 1152*f12*x2^4*x3^3 - 192*f3*x2^6*x3 + 360*f3*x2^5*x3^2 - 384*f3*x2^4*x3^3 - 480*f4*x2^6*x3 + 684*f4*x2^5*x3^2 - 1152*f4*x2^4*x3^3 + 144*x2^6*x3^2 - 96*x2^5*x3^3 + 60*f12*x2^6 + 84*f12*x2^5*x3 - 204*f12*x2^4*x3^2 - 480*f12*x2^3*x3^3 + 24*f3*x2^6 + 24*f3*x2^5*x3 - 312*f3*x2^4*x3^2 - 192*f3*x2^3*x3^3 + 60*f4*x2^6 + 132*f4*x2^5*x3 - 276*f4*x2^4*x3^2 - 480*f4*x2^3*x3^3 + 96*x2^6*x3 - 108*x2^5*x3^2 + 192*x2^4*x3^3 + 57*f12*x2^5 - 168*f12*x2^4*x3 + 204*f12*x2^3*x3^2 + 18*f3*x2^5 - 96*f3*x2^4*x3 + 120*f3*x2^3*x3^2 + 75*f4*x2^5 - 168*f4*x2^4*x3 + 276*f4*x2^3*x3^2 - 12*x2^6 - 12*x2^5*x3 + 36*x2^4*x3^2 + 96*x2^3*x3^3 - 25*f12*x2^4 + 448*f12*x2^3*x3 + 156*f12*x2^2*x3^2 - 10*f3*x2^4 + 144*f3*x2^3*x3 + 120*f3*x2^2*x3^2 - 43*f4*x2^4 + 416*f4*x2^3*x3 + 228*f4*x2^2*x3^2 - 15*x2^5 + 48*x2^4*x3 - 36*x2^3*x3^2 - 50*f12*x2^3 + 296*f12*x2^2*x3 + 12*f3*x2^3 + 96*f3*x2^2*x3 - 38*f4*x2^3 + 296*f4*x2^2*x3 + 3*x2^4 - 72*x2^3*x3 - 36*x2^2*x3^2 - 34*f12*x2^2 + 76*f12*x2*x3 - 4*f3*x2^2 + 24*f3*x2*x3 - 46*f4*x2^2 + 60*f4*x2*x3 + 6*x2^3 - 48*x2^2*x3 - 55*f12*x2 - 30*f3*x2 - 53*f4*x2 + 6*x2^2 - 12*x2*x3 - 17*f12 - 10*f3 - 19*f4 + 9*x2 + 3)$\\

\noindent $N22= -4*x2^2*(x2 - 1)^2*(180*f12*x2^4*x3^2 + 72*f3*x2^4*x3^2 + 180*f4*x2^4*x3^2 - 180*f12*x2^4*x3 + 360*f12*x2^3*x3^2 - 120*f3*x2^4*x3 + 144*f3*x2^3*x3^2 - 204*f4*x2^4*x3 + 360*f4*x2^3*x3^2 - 36*x2^4*x3^2 - 15*f12*x2^4 - 156*f12*x2^3*x3 + 180*f12*x2^2*x3^2 - 6*f3*x2^4 - 120*f3*x2^3*x3 + 72*f3*x2^2*x3^2 - 3*f4*x2^4 - 228*f4*x2^3*x3 + 180*f4*x2^2*x3^2 + 36*x2^4*x3 - 72*x2^3*x3^2 - 84*f12*x2^3 - 28*f12*x2^2*x3 - 24*f3*x2^3 - 40*f3*x2^2*x3 - 60*f4*x2^3 - 100*f4*x2^2*x3 + 3*x2^4 + 36*x2^3*x3 - 36*x2^2*x3^2 - 110*f12*x2^2 - 52*f12*x2*x3 - 36*f3*x2^2 - 40*f3*x2*x3 - 94*f4*x2^2 - 76*f4*x2*x3 + 12*x2^3 + 12*x2^2*x3 - 76*f12*x2 - 24*f3*x2 - 68*f4*x2 + 18*x2^2 + 12*x2*x3 - 19*f12 - 6*f3 - 15*f4 + 12*x2 + 3)$
\end{document}